\documentclass[pra,
,tightenlines
,showkeys
,showpacs
,twocolumn
]
{revtex4-2}

\usepackage{amsmath,amssymb}
\usepackage{bm}
\usepackage[dvipdfmx]{graphicx}
\usepackage{ascmac} 
\usepackage{fancybox}
\usepackage{float}
\usepackage{enumerate}
\usepackage{color}
\usepackage{amsthm}
\usepackage{mathrsfs}
\usepackage{comment}
\allowdisplaybreaks[1]
\usepackage{comment}
\newtheorem*{th.}{Theorem}

\newcommand{\ket}[1]{|#1\rangle}
\newcommand{\bra}[1]{\langle #1|}
\newcommand{\braket}[2]{\langle#1|#2\rangle}

\usepackage{ulem}

\begin{document}
	\title{Mitigating Detuning-Induced Systematic Errors in Entanglement-Enhanced Metrology}
    \author{Shingo Kukita}
    \email{kukita@nda.ac.jp}
    \affiliation{National Defense Academy of Japan}
    \author{Yuichiro Matsuzaki}
    \email{ymatsuzaki872@g.chuo-u.ac.jp}
    \affiliation{Chuo University}    
    \begin{abstract}
        Quantum sensing leverages non-classical resources to enhance precision. In particular, Greenberger–Horne–Zeilinger (GHZ) states can, in principle, attain the Heisenberg limit that surpasses the standard quantum limit. While many studies have examined how open-system noise—typically modeled with Lindblad master equations—degrades GHZ-based metrology, coherent control imperfections during state preparation and readout have received less attention. Here, we analyze the effect of detuning between actual and nominal spin frequencies in a GHZ-state preparation scheme employing a frequency selective pulse. We show that detuning induces coherent, systematic error that prevents GHZ sensing from reaching the Heisenberg limit. To mitigate this effect, we design a composite-pulse protocol that compensates for detuning-induced errors and improves the sensitivity under the effect of coherent error. 
        \end{abstract}
    \maketitle
	\section{introduction}
    
    Precise measurement plays a central role in modern technology~\cite{degen2017quantum,barry2020sensitivity}.
    Magnetic-field measurements, in particular, find broad application in physics~\cite{montenegro2025quantum}, materials science~\cite{sun2013recent}, chemistry~\cite{frasco2009semiconductor}, and biology~\cite{aslam2023quantum}.
    A typical magnetometer employs $N$-spin probes~\cite{wolf2015subpicotesla}.
    The applied magnetic field modifies the spin Hamiltonian;
    after an exposure time, the spins accumulate a field-dependent relative phase ~\cite{maze2008nanoscale,balasubramanian2008nanoscale,taylor2008high}.
    Reading out this phase, one can estimate the field strength with suitable protocols.
    
    When estimating the field strength with $N$ independent, identically prepared spins, the standard deviation typically scales as $\mathcal{O}(N^{-1/2})$, known as the standard quantum limit (SQL)~\cite{huelga1997improvement,giovannetti2004quantum,huang2024entanglement}.
    In contrast, if the $N$ spins are entangled into a Greenberger--Horne--Zeilinger (GHZ) state, the standard deviation can scale as $\mathcal{O}(N^{-1})$, the Heisenberg limit~\cite{huelga1997improvement}.
    GHZ-based sensing is powerful in the ideal, noiseless limit, but its performance degrades under stochastic noise that induces decoherence during exposure to the target field.
    In particular, under time-homogeneous (Markovian) noise, GHZ protocols cannot surpass the SQL~\cite{huelga1997improvement,shaji2007qubit}.
    When the noise is time-inhomogeneous so that the leading-order decoherence is quadratic in the exposure time (short-time Zeno regime),
    an optimally timed GHZ protocol can beat the SQL and achieve a sensitivity scaling of $\mathcal{O}(N^{-3/4})$, which is called the Zeno limit~\cite{matsuzaki2011magnetic,chin2012quantum,long2022entanglement}.

    The impact of diverse noise processes on GHZ sensing has been extensively analyzed.
    By contrast, the effect of control imperfections during GHZ-state preparation and readout on the achievable sensitivity has received far less attention.
    Such imperfections typically manifest themselves as systematic (bias) errors.
    Unlike stochastic noise, which primarily inflates estimator variance and slows the decay of standard deviations, systematic errors can bias the estimator and prevent its convergence to the true value\cite{sugiyama2015precision,takeuchi2019quantum}.
    It is therefore crucial to quantify and mitigate their impact.
    Although systematic errors are ubiquitous in experiments, a broadly applicable mitigation framework remains elusive; most prior studies treat decoherence~\cite{pang2016protecting,shimada2021quantum,yamamoto2022error} rather than control errors.
    
    We consider the experimentally accessible GHZ-sensing protocol used in Refs.~\cite{jones2009magnetic,PhysRevLett.115.170801}.
    In this approach, a global control pulse is applied simultaneously to $N$ spins with frequency selectivity to prepare the GHZ probe state.
    Fluctuations in parameters of apparatuses and pulses induce coherent mis-rotations,
    so each spin deviates from its target state;
    as a result, the prepared state approaches a superposition of spin-coherent product states rather than an ideal GHZ state.
    Ref.~\cite{PhysRevLett.115.170801} further shows that for a certain class of pulse imperfections, e.g., uniform amplitude miscalibration, the protocol can still exhibit $O(N^{-1})$ scaling, and under time-inhomogeneous dephasing, it can attain a sensitivity scaling of $O(N^{-3/4})$.
    However, Ref.~\cite{PhysRevLett.115.170801} does not analyze the effect of pulse-frequency deviations, which is called detuning.
    As shown below, detuning introduces a coherent, systematic bias into the above GHZ-sensing protocol and, as the number of spins increases, significantly hinders the convergence of the estimator to the true value.

    We quantify the impact of detuning on GHZ-based sensing.
    Moreover, to mitigate this effect, we propose a compensation method based on composite pulses, originally developed in nuclear magnetic resonance (NMR)~\cite{counsell1985analytical,levitt1986composite,claridge2016high,Levitt2008}. Composite pulses replace a single target rotation with an appropriately designed sequence of rotations that cancels leading-order control errors.
    They are widely used for robust control~\cite{PhysRevA.77.052334,Ball_2021,PhysRevX.8.021058,Zanon-Willette_2018,said2009robust}.
    Recently, an application of composite pulses to quantum sensing was also proposed~\cite{lancaster2025quantum}.
    In this paper, we introduce a composite-pulse sequence tailored to GHZ-state preparation and evaluate its performance.

    The remainder of this paper is organized as follows:
    Section~\ref{sec:2} briefly reviews quantum sensing and the effects of systematic errors.
    In Sec.~\ref{sec:3}, we introduce the sensing scheme 
    used
    in Ref.~\cite{jones2009magnetic,PhysRevLett.115.170801}, and evaluate how detuning affects its performance.
    Furthermore, we provide a composite pulse sequence that compensates for the effect of detuning.
    In Sec.~\ref{sec:4}, the performance of our proposal is numerically evaluated.
    Section~\ref{sec:5} is devoted to the summary of this paper.

    \section{review of quantum sensing and effects of systematic errors}
    \label{sec:2}

    Quantum sensing is formally treated as a parameter estimation problem, where a parameter $\Omega$ encoded in a quantum state $\rho_\Omega$ is estimated via measurements.
    Let $\tilde{\Omega}$ be an estimator constructed from measurement outcomes.
    We denote the statistical expectation value over the measurement results by $\langle \bullet \rangle$.
    The estimation precision is quantified by the mean squared error (MSE), defined as
    \begin{equation}
    \mathrm{MSE}(\tilde{\Omega}) = \langle (\tilde{\Omega} - \Omega)^2 \rangle.
    \end{equation}
    The MSE can be decomposed into the statistical variance and the systematic bias:
    \begin{equation}
    \mathrm{MSE}(\tilde{\Omega}) = \mathrm{Var}(\tilde{\Omega}) + [b(\Omega)]^2,
    \label{eq:MSE_decomp}
    \end{equation}
    where $\mathrm{Var}(\tilde{\Omega}) = \langle (\tilde{\Omega} - \langle \tilde{\Omega} \rangle)^2 \rangle$ represents the fluctuation around the mean, and $b(\Omega) = \langle \tilde{\Omega} \rangle - \Omega$ represents the bias~\cite{kay1993fundamentals}.

    For unbiased estimators where $b(\Omega)=0$, the MSE reduces to the variance.
    The ultimate precision limit of any estimator $\tilde{\Omega}$ is governed by the Quantum Cramér--Rao Bound (QCRB):
    \begin{equation}
    \mathrm{Var}(\tilde{\Omega}) \geq \frac{1}{M F_Q(\rho_\Omega)},
    \label{eq:QCRB}
    \end{equation}
    where $M$ is the number of repetitions and $F_Q(\rho_\Omega)$ is the Quantum Fisher Information (QFI), which depends solely on the probe state and its evolution~\cite{helstrom1976quantum}.
    When a specific measurement (POVM) is performed, yielding outcomes $s$ with probability $P(s|\Omega)$,
    the precision is further bounded by the classical Cramér--Rao bound (CCRB):
    \begin{equation}
    \mathrm{Var}(\tilde{\Omega}) \geq \frac{1}{M F(\Omega)} \geq \frac{1}{M F_Q(\rho_\Omega)},
    \label{eq:CCRB}
    \end{equation}
    where $F(\Omega) = \sum_s P(s|\Omega) (\partial_\Omega \ln P(s|\Omega))^2$ is the classical Fisher information (see Ref.~\cite{paris2009quantum} for a comprehensive review).

    A rigorous approach to quantum metrology involves simultaneously optimizing the initial probe state, the measurement basis, and the estimator to saturate the QCRB. However, in this work, we do not delve into such global optimization. Instead, we focus on a specific, well-established class of sensing protocols---utilizing GHZ states.
    In the ideal noiseless limit, a measurement saturating the QCRB and achieving the Heisenberg limit scaling $\mathcal{O}(N^{-1})$ for $N$ spins has been found.
    Our primary objective is to address a critical practical limitation of this scheme:
    its vulnerability to control errors.
    As indicated by Eq.~\eqref{eq:MSE_decomp}, systematic errors ($b(\Omega) \neq 0$) introduce a constant bias term to the MSE, which cannot be suppressed by increasing $M$.
    
    As a concrete example, we consider spin-based magnetometry, in which a magnetic field is encoded as a phase on spins (or qubits) during an exposure period.
    The state space of a spin is spanned by two vectors $\{\ket{g},\ket{e}\}$.
    We introduce the Pauli matrices $(\sigma_{0},\sigma_{x},\sigma_{y},\sigma_{z})$ through their actions as
    \begin{align}
        \sigma_{0}\ket{g}=&\ket{g},&\sigma_{0}\ket{e}=&\ket{e},\nonumber\\
        \sigma_{x}\ket{g}=&\ket{e},&\sigma_{x}\ket{e}=&\ket{g},\nonumber\\
        \sigma_{y}\ket{g}=&i\ket{e},&\sigma_{y}\ket{e}=&-i\ket{g},\nonumber\\
        \sigma_{z}\ket{g}=&-\ket{g},&\sigma_{z}\ket{e}=&\ket{e},
    \end{align}
    where $\sigma_{0}$ is the two-dimensional identity matrix.
    We assume that the target magnetic field is directed to the $z$-axis and its strength is $\Omega$ with appropriate normalizations.
    The Hamiltonian of a spin exposed to the magnetic field is given as
    \begin{equation}
        H=\frac{\Omega}{2}\sigma_{z}.
        \label{eq:mag_single}
    \end{equation}
    
    We initially prepare the spin in the superposition state $\ket{+_{x}}:=\left(\ket{g}+\ket{e}\right)/\sqrt{2}$.
    After being exposed to the magnetic field, the state becomes
    \begin{equation}
        \ket{\psi(\Omega)}=\frac{e^{i\Omega t/2}}{\sqrt{2}}\left(\ket{g}+e^{-i\Omega t}\ket{e}\right),
    \end{equation}
    where $t$ is the exposure time.
    Then, we perform the projective measurement ${\cal P}_{y\pm}=(\sigma_{0}\pm \sigma_{y})/2$.
	The probability of obtaining the outcome ``$y+$" is given as
	\begin{equation}
	P_{y{+}}=\bra{\psi(\Omega)}{\cal P}_{y+}\ket{\psi(\Omega)}=\frac{1}{2}+\frac{\sin(\Omega t)}{2}\approx \frac{1}{2}+\frac{\Omega t}{2},
	\end{equation} 
    where we assume $\Omega t \ll 1$.
    As $P_{y{+}}$ has information on the field strength $\Omega$,
    we can estimate $\Omega$ through $P_{y{+}}$ by repeating the above process.

    We assign the value $1$ to the outcome $y+$ while assigning $0$ to $y-$.
    Let $s_{j}\in \{0,1\}$ denote the outcome in the $j$th experiment. 
    Repeat the experiment $M$ times and calculate the following quantity:
    \begin{equation}
        S=\sum_{j=1}^{M}s_{j}/M.
    \end{equation}
    If $M$ is sufficiently large, this value converges to $P_{y+}$.
    We can define an estimator $\tilde{\Omega}$ of the field strength $\Omega$ as
    \begin{equation}
        \tilde{\Omega}=\frac{2 S -1}{t},
        \label{eq:estimator}
    \end{equation}
    and $\tilde{\Omega}\rightarrow \Omega$ for large $M$.
    The root-mean-square error (RMSE) of $\tilde{\Omega}$ from the true value $\Omega$ is calculated as
    \begin{equation}
        \sqrt{\mathrm{MSE}(\tilde{\Omega})}=\sqrt{\langle (\tilde{\Omega}-\Omega)^{2} \rangle} = \frac{2}{t}\sqrt{\frac{P_{y+}\left(1-P_{y+}\right)}{M}}.
        \label{eq:stddev_iid}
    \end{equation}
    By increasing the number of experiments $M$, which here corresponds to the number of spins utilized in estimation, the standard deviation decreases, following the scaling law ${\cal O}(M^{-1/2})$.
    As demonstrated below, when the GHZ sensing scheme with $N$ spins is performed, the exposure time $t$ in Eq.~(\ref{eq:stddev_iid}) is effectively replaced by $Nt$. This yields the scaling law ${\cal O}(N^{-1})$, corresponding to the Heisenberg limit.

    Let us explain the generalized setup \cite{sugiyama2015precision,takeuchi2019quantum}. We consider $M$ experiments, each of which yields an outcome of $1$ with probability $P_{1}$ and $0$ with probability $1-P_{1}$.
    Assume that the probability $P_{1}$ is described by
    \begin{equation}
        P_{1}=x+y\Omega,
    \end{equation}
    as a function of $\Omega$, the parameter to be estimated.
    Then defining the estimator $\tilde{\Omega}$ in the same way as in the above example, we obtain
    \begin{equation}
        \langle \tilde{\Omega} \rangle =\Omega,~~\sqrt{\mathrm{MSE}(\tilde{\Omega})} = \frac{1}{y}\sqrt{\frac{P_{1}\left(1-P_{1}\right)}{M}}.
        \label{eq:stddev_general}
    \end{equation}
    However, the situation changes when the process is subject to unknown systematic errors.
    In this case, the actual probability $P'_{1}$ is 
    \begin{equation}
        P'_{1}=x'+y'\Omega,
    \end{equation}
    where $x'$ ($y'$) differs from $x$ ($y$) while we assume that the outcomes are generated by the ideal probability $P_{1}$.
    Hence, the estimator (\ref{eq:estimator}) has an incorrect average:
    \begin{equation}
        \langle \tilde{\Omega} \rangle =\frac{y'\Omega+x'-x}{y}.
    \end{equation}
    Note that we now take the statistical average according to the actual probability $P'_{1}$.
    Consequently, the RMSE behaves as
    \begin{equation}
        \sqrt{\mathrm{MSE}(\tilde{\Omega})}=\frac{1}{y}\sqrt{\frac{P'_{1}\left(1-P'_{1}\right)}{M}+\left(\Omega \Delta y +\Delta x\right)^{2}},
    \end{equation}
    where $\Delta x=x'-x$ and $\Delta y=y'-y$.
    We observe that this does not converge to zero at the limit of $M\rightarrow \infty$:
    \begin{equation}
        \lim_{M\to \infty}\sqrt{\mathrm{MSE}(\tilde{\Omega})}=\frac{1}{y}|\Omega \Delta y +\Delta x|=|b\left(\Omega\right)|.
    \end{equation}

    \section{error-robust GHZ sensing}
    \label{sec:3}
    \subsection{ideal case}

    As a starting point, we examine the conventional GHZ sensing scheme. This scheme achieves the Heisenberg limit by employing the GHZ state as the initial state.
    The GHZ state of $N$ spins is defined as
    \begin{equation}
        \ket{{\rm GHZ}}=\frac{1}{\sqrt{2}}\left(\ket{\underbrace{gg\cdots g}_{N}}+\ket{\underbrace{ee\cdots e}_{N}}\right).
    \end{equation}
    When the spins comprising this state are subjected to a uniform magnetic field, as given in Eq.~(\ref{eq:mag_single}), the state evolves into the following form:
    \begin{equation}
        \ket{{\rm GHZ'}} :=\frac{e^{iN\Omega t/2}}{\sqrt{2}}\left(\ket{gg\cdots g}+e^{-iN\Omega t}\ket{ee\cdots e}\right).
    \end{equation}
    We measure this state by the projection operator ${\cal P}_{1}=\ket{{\rm GHZ}_{y}}\bra{{\rm GHZ}_{y}}$ and ${\cal P}_{0}=1-{\cal P}_{1}$, where $\ket{{\rm GHZ}_{y}}=\left(\ket{gg\cdots g}+i\ket{ee\cdots e}\right)/\sqrt{2}$.
    The probability of obtaining the outcome $1$ is given as
    \begin{equation}
        P_{1}=\left|\braket{{\rm GHZ}_{y}}{{\rm GHZ}'}\right|^{2}\approx\frac{1}{2}+\frac{N\Omega t}{2}
    \end{equation}
    where we assume $N \Omega t\ll 1$.
    According to Eq.~(\ref{eq:stddev_general}), we can achieve the Heisenberg limit ${\cal O}(N^{-1})$.
    It is worth noting that this measurement and estimator saturate the QCRB in the ideal noiseless limit, provided that the accumulated phase is small.
    
    In the conventional GHZ sensing scheme, one must perform a measurement with respect to the $\ket{{\rm GHZ}_{y}}\bra{{\rm GHZ}_{y}}$ basis, which is typically challenging to realize experimentally. To overcome this, Ref.~\cite{PhysRevLett.115.170801} proposed a practical GHZ sensing scheme that circumvents the need for direct $\ket{{\rm GHZ}_{y}}\bra{{\rm GHZ}_{y}}$ measurements. Here, the system consists of a controllable spin (c) that is collectively coupled to $N$ memory spins (m).
	The Hamiltonian of the system is
    \begin{align}
	H(t)=&\omega_{c}\frac{\sigma_{z}^{(c)}}{2}+\omega_{m}\frac{J_{z}^{(m)}}{2}+g\frac{\sigma_{z}^{(c)}J_{z}^{(m)}}{4}\nonumber\\
    &+ \lambda \frac{J^{(m)}_{x}}{2}\cos(\omega t-\phi).
	\label{eq:hamiltonian}
    \end{align}
    In this notation, the superscripts in parentheses specify the target spin of each operator: $(c)$ corresponds to the controllable spin, whereas $(m)$ corresponds to the memory spins.
    The collective spin operators $J_{z,x}^{(m)}$ are defined by $J_{x,y,z}^{(m)}=\sum_{i=1}^{N}\sigma^{(m,i)}_{x,y,z}$, where $\sigma^{(m,i)}_{x,y,z}$ acts on the $i$-th spin among the memory ones.
    The first two terms in Eq.~(\ref{eq:hamiltonian}) correspond to the spin frequencies, the third term describes the coupling between the controllable spin and the memory spins, and the last term represents the pulse applied to the memory spins.
    The pulse strength $\lambda$ is set to unity, and other parameters such as $\omega_{m,c}$, $g$, and time $t$ are normalized with respect to this value. We assume that the parameters of $(\omega,\phi)$ in the last term are controllable. The resonance frequencies of the memory spins are assumed to be identical.
    
    We take the initial state to be $\ket{g}_{c}\otimes \ket{g\cdots g}$, where the controllable spin is written on the leftmost side. A resonant pulse is first applied to the controllable spin, resulting in the following state:
	\begin{align}
	\ket{\Psi_{0}}=&\frac{1}{\sqrt{2}}\left(\ket{g}_{c}+\ket{e}_{c}\right)\otimes \ket{\underbrace{ggg\cdots g}_{N}}\nonumber\\
    =&\frac{1}{\sqrt{2}}\ket{g}_{c}\otimes\ket{g\cdots g}+\frac{1}{\sqrt{2}}\ket{e}_{c}\otimes \ket{g\cdots g}.
	\label{eq:initial}
	\end{align}
    We assume that this pulse is ideal, introducing no error, and that it does not influence the memory spins because the frequency detuning satisfies $|\omega_{c}-\omega_{m}|\gg 1$.

    Next, we apply a pulse to the memory spins. To analyze their behavior, we rewrite Hamiltonian~(\ref{eq:hamiltonian}) in the rotating frame defined by the first three terms:
    \begin{align}
        \tilde{H}(t)=&\ket{g}_{c}\bra{g}\otimes \tilde{H}_{-}(\omega,t)+\ket{e}_{c}\bra{e}\otimes \tilde{H}_{+}(\omega,t),\nonumber\\
        \tilde{H}_{\pm}(\omega,t):=&\frac{J^{(m)}_{x}\cos (\omega_{\pm}t)-J^{(m)}_{y}\sin (\omega_{\pm}t)}{2}\cos(\omega t-\phi),
        \label{eq:decomposed_hamiltonian}
    \end{align}
    where $\omega_{\pm}=\omega_{m}\pm g/2$.
    This Hamiltonian does not induce transitions of the controllable spin.

    Under the condition $\omega_{m}, g \gg 1$, the partial Hamiltonians $H_{\pm}$ during this pulse can be approximated as
    \begin{align}
        \tilde{H}_{+}(\omega_{+},t)\approx&~\tilde{H}_{\rm res}:=\frac{1}{4}\left(J^{(m)}_{x}\cos\phi +J^{(m)}_{y} \sin\phi\right),\nonumber\\
        \tilde{H}_{-}(\omega_{+},t)\approx&~0,
        \label{eq:quater}
    \end{align}
    where we perform the rotating wave approximation.
    The validity of this approximation, including the impact of counter-rotating terms, is discussed in Appendix~\ref{app:RWA}.
    Setting $\phi=-\pi/2$ and (normalized) pulse duration $2\pi$,
    we obtain the following state after the pulse:
	\begin{equation}
	\ket{\Psi_{1}}=\frac{1}{\sqrt{2}}\left(\ket{g}_{c}\otimes\ket{g\cdots g}+\ket{e}_{c}\otimes\ket{e\cdots e}\right).
	\end{equation}
    It should be noted that the basis states $\ket{e}$ and $\ket{g}$ are now defined in the rotating frame. For simplicity, we omit the explicit notation, as this distinction does not influence the result.
    
    Next, we subject the state to the target magnetic field. The following additional Hamiltonian describes the effect of this field:
    \begin{equation}
        H_{\rm mag}=\frac{\Omega}{2}J^{(m)}_{z}.
    \end{equation}
    The state evolves to
	\begin{equation}
	\ket{\Psi_{2}}=\frac{1}{\sqrt{2}}\left(e^{i \frac{N\Omega t}{2}}\ket{g}_{c}\otimes\ket{g\cdots g}+e^{-i \frac{N\Omega t}{2}}\ket{e}_{c}\otimes\ket{e\cdots e}\right).
	\end{equation}
	We then apply a pulse with $\omega=\omega_{-}$.
    With this choice of frequency, the partial Hamiltonians $H_{\pm}$ take the following form:
    \begin{equation}
        \tilde{H}_{+}(\omega_{-},t)\approx~0,~~\tilde{H}_{-}(\omega_{-},t)\approx~\tilde{H}_{\rm res}.        
    \end{equation}
    When we set $\phi=-\pi/2$ and choose a pulse duration of $2\pi$, the resulting state is
	\begin{align}
    \ket{\Psi_{3}}=&
	\frac{1}{\sqrt{2}}\left(e^{i \frac{N\Omega t}{2}}\ket{g}_{c}\otimes\ket{e\cdots e}+e^{-i \frac{N\Omega t}{2}}\ket{e}_{c}\otimes\ket{e\cdots e}\right)\nonumber\\
	=&\frac{1}{\sqrt{2}}\left(e^{i \frac{N\Omega t}{2}}\ket{g}_{c}+e^{-i \frac{N\Omega t}{2}}\ket{e}_{c}\right)\otimes\ket{e\cdots e}.
	\end{align}
    In the above protocol, we assume that the target magnetic field is decoupled during both the state preparation and readout processes.
    
	We measure the controllable spin using the basis of $\ket{\pm_{y}}=(\ket{e}_{c}\pm i\ket{g}_{c})/\sqrt{2}$. The probability of obtaining $\ket{+_{y}}$, which we define as outcome $1$, is expressed as
	\begin{equation}
	P_{+y}\approx\frac{1}{2}+\frac{N\Omega t}{2},
	\end{equation}
	where we assume $N\Omega t \ll 1$.
    Reference~\cite{PhysRevLett.115.170801} showed that this sensing protocol can achieve the Heisenberg limit even in the presence of certain pulse errors in the memory spins. However, as we will demonstrate, detuning errors introduced below prevent the scheme from attaining the Heisenberg limit.
	
	\subsection{sensing with detuning error}

    In the above derivation, we assumed identical resonance frequencies for all memory spins. In reality, each frequency may exhibit a small deviation from the ideal value, expressed as $\omega_{m}+\delta_{i}$, with $i$ labeling the spin. We call this deviation detuning, and the corresponding Hamiltonian is given as
   \begin{align}
        \tilde{H}'(t)=&\ket{g}_{c}\bra{g}\otimes \tilde{H}'_{-}(\omega,t)+\ket{e}_{c}\bra{e}\otimes \tilde{H}'_{+}(\omega,t),\nonumber\\
        \tilde{H}'_{\pm}(\omega,t):=&\frac{J^{(m)}_{x}\cos (\omega_{\pm}t)-J^{(m)}_{y}\sin (\omega_{\pm}t)}{2}\cos(\omega t-\phi)\nonumber\\
        &+\sum_{i=1}^{N}\delta_{i}\frac{\sigma^{(m,i)}_{z}}{2},
        \label{eq:erroneous2}
    \end{align}
    in the rotating frame.
    Consequently, when we tune $\omega=\omega_{\pm}$, the partial Hamiltonians $\tilde{H}'_{\pm}$ are approximated as
    \begin{align}
        \tilde{H}'_{+}(\omega_{+},t)=\tilde{H}'_{-}(\omega_{-},t)\approx&~\tilde{H}_{\rm res}+\sum_{i=1}^{N}\delta_{i}\frac{\sigma^{(m,i)}_{z}}{2},\nonumber\\
        \tilde{H}_{-}(\omega_{+},t)=\tilde{H}_{+}(\omega_{-},t)\approx&~\sum_{i=1}^{N}\delta_{i}\frac{\sigma^{(m,i)}_{z}}{2}.    
        \label{eq:erroneous_hamiltonian}
    \end{align}
    We assume that detuning is unknown but constant throughout the whole process, while varying across different spins.
    
    We reconsider the sensing protocol in the presence of detuning. Beginning with the state~(\ref{eq:initial}), a pulse characterized by $\omega=\omega_{+}$, $\phi=-\pi/2$, and duration $2\pi$ is applied, leading to the following state in the rotating frame:
	\begin{align}
	\ket{\Psi_{1,\delta}}=&\frac{1}{\sqrt{2}}\left(e^{i\pi \Delta}\ket{g}_{c}\otimes\ket{g\cdots g}+\ket{e}_{c}\otimes\ket{\psi_{1}\cdots \psi_{N}}\right),
	\end{align}
	where $\Delta=\sum_{i=1}^{N}\delta_{i}$ and
	\begin{equation}
	\ket{\psi_{i}}=\ket{e}-2 i\delta_{i}\ket{g}+{\cal O}\left(\delta_{i}^{2}\right).
	\end{equation}

    We next expose this state to the target magnetic field. In this process, detuning also affects the dynamics. The corresponding Hamiltonian is given by
    \begin{equation}
        H_{\rm mag}=\frac{\Omega}{2}J^{(m)}_{z}+\sum_{i=1}^{N}\delta_{i}\frac{\sigma^{(m,i)}_{z}}{2}.
    \end{equation}
    Under this Hamiltonian, the time evolution of the state is described as
	\begin{align}
	\ket{\Psi_{2,\delta}}=&\frac{1}{\sqrt{2}}e^{i \frac{(N\Omega+\Delta) t}{2}}e^{i\pi \Delta}\ket{g}_{c}\otimes\ket{g\cdots g}\nonumber\\
    &+\frac{1}{\sqrt{2}}e^{-i \frac{(N\Omega+\Delta) t}{2}}\ket{e}_{c}\otimes\ket{\psi'_{1}\cdots \psi'_{N}},
	\end{align}
	where
	\begin{equation}
	\ket{\psi'_{i}}=\ket{e}-ie^{i (\Omega+\delta_{i}) t}\delta_{i}\ket{g}+{\cal O}\left({\delta_{i}}^{2}\right).
	\end{equation}
    We then apply a pulse with $\omega=\omega_{-}$, $\phi=-\pi/2$, and duration $2\pi$, after which the state evolves into $\ket{\Psi_{3,\delta}}$:
	\begin{align}
	\ket{\Psi_{3,\delta}}=&\frac{1}{\sqrt{2}}e^{i \frac{(N\Omega+\Delta) t}{2}}e^{i\pi \Delta}\ket{g}_{c}\otimes\ket{\psi_{1}\cdots \psi_{N}}\nonumber\\
    &+\frac{1}{\sqrt{2}}e^{-i \frac{(N\Omega+\Delta) t}{2}}e^{-i\pi \Delta}\ket{e}_{c}\otimes\ket{\psi''_{1}\cdots \psi''_{N}},
	\end{align}
	where
	\begin{equation}
	\ket{\psi''_{i}}=\ket{e}-i e^{i (\Omega+\delta_{i}) t}e^{2i \pi \delta_{i}}\delta_{i}\ket{g}+{\cal O}(\delta_{i}^{2}).
	\end{equation}	
	We measure the controllable spin by $\{\ket{\pm_{y}}=\left(\ket{e}_{c}\pm i \ket{g}_{c}\right)/\sqrt{2}\}$.
	The probability of obtaining $\ket{+_{y}}$ is
	\begin{equation}
	P_{+y}\approx\frac{1}{2}\left(1+(2\pi+t)\Delta\right)+\frac{N\Omega t}{2}+{\cal O}(\delta^{2}),
    \label{eq:sys_err}
    \end{equation}
    where $\delta^{2}$ denotes all possible products from $\delta_{i}$'s, such as $\delta^{2}_{i}$ and $\delta_{i}\delta_{j}$.
    Also, we use $N\Omega t \ll 1$ and $1-|\braket{\psi_{i}}{\psi''_{i}}|={\cal O}(\delta_{i}^{2})$ in the derivation.
    
    A systematic error arises from $\Delta$. The impact of this error is determined by the distribution of the $\delta_{i}$’s. For uniform detuning $\delta_{i}=\delta$, then the statistical average and the RMSE of the estimator $\tilde{\Omega}$ are given by
    \begin{align}
        \langle \tilde{\Omega}\rangle\approx&~\Omega + \left(\frac{2\pi}{t}+1\right)\delta,\nonumber\\
        \lim_{M\to \infty}\sqrt{\mathrm{MSE}(\tilde{\Omega})}=&~\left(\frac{2\pi}{t}+1\right)|\delta |.
        \label{eq:mse_error}
    \end{align}
    These results indicate that the estimator fails to converge to the true value $\Omega$, which represents a significant limitation to achieving accurate sensing.
    
    Note that the dynamics $\ket{g}\rightarrow\ket{\psi_{i}}$ under the effect of detuning does not degrade sensing precision up to ${\cal O}(\delta)$, because $1-|\braket{\psi_{i}}{\psi''_{i}}|={\cal O}(\delta_{i}^{2})$. By contrast, the effect of detuning in the other processes, such as $\ket{g}\rightarrow e^{i\pi\Delta}\ket{g}$ and $\ket{\psi'_{i}}\rightarrow e^{-i\pi\Delta}\ket{\psi''_{i}}$, have a more critical influence.
    Therefore, compensating for the contributions from the latter processes is rather essential for improving sensing accuracy.

    A key assumption in our protocol is that the target magnetic field remains decoupled during the state preparation and readout processes.
    If the target magnetic field were always present simultaneously with detuning, we could not distinguish the target magnetic field from detuning.

	\subsection{composite pulse sequence}

    Here, we focus on the state preparation process that involves the $\omega_{+}$ pulse and modify the pulse sequence. We assume that the pulse strength $\lambda$ is limited by physical constraints. This assumption is necessary; otherwise, the crucial condition $g \gg 1 (=\lambda)$ would no longer hold.
    We also fix the pulse strength throughout the protocol to avoid errors from its variation. Relaxing this constraint would allow a simpler and more efficient protocol than that described below (Appendix~\ref{sec:app}).

    First, note that if we apply only $\omega_{+}$ pulses, errors in the dynamics of $\ket{g}_{c}\otimes\cdots$ cannot be compensated. The dynamics of this component depends on the pulse duration and not on the pulse details. Therefore, it is necessary to switch from $\omega{\pm}$ to $\omega_{\mp}$ at some point in the pulse sequence. The condition to be satisfied when exciting the ground state tensored with $\ket{e}_{c}$ is
	\begin{align}
	\ket{e}_{c}\otimes\ket{g\cdots g}~\rightarrow &~\ket{e}_{c}\otimes\ket{\xi_{1}\xi_{2}\cdots\xi_{n}},\nonumber\\
    \ket{\xi_{i}}~=&~\ket{e}-2 i\alpha\delta_{i}\ket{g}+{\cal O}(\delta^{2}),\nonumber\\
	\ket{g}_{c}\otimes\ket{g\cdots g}~\rightarrow &~e^{-i \frac{\delta_{i}t}{2}}\ket{g}_{c}\otimes\ket{gg\cdots g}+{\cal O}(\delta^{2}),
	\end{align}
    Here, $t$ denotes the duration of exposure to the magnetic field (not the duration of pulses), and $\alpha$ is an arbitrary constant of order ${\cal O}(1)$. The vector $\ket{\xi_{i}}$ is not necessarily equal to $\ket{\psi_{i}}$. To compensate for the effect of detuning during the exposure process, it is necessary to include the phase factor $e^{-i\delta_{i}t/2}$.
    
	By straightforward but lengthy calculations, we can show that the following pulse sequence satisfies the above conditions with $\alpha=1$:
	\begin{enumerate}[I.]
	\item $\omega=\omega_{-}$, $\phi=\phi_{1}$, $s/2=\arcsin\left((\pi+t/2)/8\right)$,
	\item $\omega=\omega_{-}$, $\phi=\phi_{1}+\pi$, $s/2=2\arcsin\left((\pi+t/2)/8\right)$,
	\item $\omega=\omega_{-}$, $\phi=\phi_{1}$, $s/2=\arcsin\left((\pi+t/2)/8\right)$,
	\item $\omega=\omega_{+}$, $\phi=-\pi/2$, $s/2=\pi$
	\item $\omega=\omega_{-}$, $\phi=\phi_{1}$, $s/2=\arcsin\left((\pi+t/2)/8\right)$,
	\item $\omega=\omega_{-}$, $\phi=\phi_{1}+\pi$, $s/2=2\arcsin\left((\pi+t/2)/8\right)$,
	\item $\omega=\omega_{-}$, $\phi=\phi_{1}$, $s/2=\arcsin\left((\pi+t/2)/8\right)$,
	\end{enumerate} 
	where $\phi_{1}$ is an arbitrary pulse phase, $s$ represents the duration of each pulse.
    The factor $1/2$ of $s$ originates from the factor $1/4$ of the pulse strength in $\tilde{H}_{\rm res}$.
	The fourth $\omega_{+}$-pulse is sandwiched between two identical composite $\omega_{-}$-pulses, both of which implement the identity operation when $\delta_{i}=0$.
    These $\omega_{-}$-pulses are applied to ``accumulate" the effects of detuning enough to cancel out those in the other processes.
    The exposure duration $t$ is restricted to be less than $2(8-\pi)$ owing to the property of the $\arcsin$ function. Here, $t$ is normalized by the pulse strength. Consequently, weakening the pulse strength increases the actual exposure time, which may improve the estimation of $\Omega$. However, this also amplifies the effect of detuning during the exposure.
    
	The state after this operation is
	\begin{align}
	\ket{\Psi^{\rm CP}_{1,\delta}}=&\frac{1}{\sqrt{2}}~e^{-i \frac{\Delta t}{2}}\ket{g}_{c}\otimes\ket{g\cdots g}\nonumber\\
    &+\frac{1}{\sqrt{2}}\ket{e}_{c}\otimes\ket{\psi_{1}\cdots \psi_{N}}+{\cal O}(\delta^{2}),
	\end{align}
	Exposing this state to a target magnetic field
    with detuning, we obtain the following state:
	\begin{align}
	\ket{\Psi^{\rm CP}_{2,\delta}}=&\frac{1}{\sqrt{2}}e^{i \frac{N\Omega t}{2}}\ket{g}_{c}\otimes\ket{g\cdots g}\nonumber\\
    &+\frac{1}{\sqrt{2}}e^{-i \frac{(N\Omega+\Delta) t}{2}}\ket{e}_{c}\otimes\ket{\psi'_{1}\cdots \psi'_{N}}+{\cal O}(\delta^{2}),
	\end{align}
	Furthermore, the following similar pulse sequence,
	\begin{enumerate}[I.]
	\item $\omega=\omega_{+}$, $\phi=\phi_{1}$, $s/2=\arcsin\left((\pi+t/2)/8\right)$,
	\item $\omega=\omega_{+}$, $\phi=\phi_{1}+\pi$, $s/2=2\arcsin\left((\pi+t/2)/8\right)$,
	\item $\omega=\omega_{+}$, $\phi=\phi_{1}$, $s/2=\arcsin\left((\pi+t/2)/8\right)$,
	\item $\omega=\omega_{-}$, $\phi=-\pi/2$, $s/2=\pi$,
	\item $\omega=\omega_{+}$, $\phi=\phi_{1}$, $s/2=\arcsin\left((\pi+t/2)/8\right)$,
	\item $\omega=\omega_{+}$, $\phi=\phi_{1}+\pi$, $s/2=2\arcsin\left((\pi+t/2)/8\right)$,
	\item $\omega=\omega_{+}$, $\phi=\phi_{1}$, $s/2=\arcsin\left((\pi+t/2)/8\right)$,
	\end{enumerate} 
	satisfies
	\begin{align}
    \ket{e}_{c}\otimes\ket{\psi'_{1}\psi'_{2}\cdots\psi'_{N}}~\rightarrow&~e^{i\frac{\delta_{i}t}{2}}\ket{e}_{c}\otimes\ket{\psi''_{1}\cdots\psi''_{N}}+{\cal O}(\delta^{2}),\nonumber\\
	\ket{g}_{c}\otimes\ket{gg\cdots g}~\rightarrow &~\ket{g}_{c}\otimes\ket{\psi_{1}\psi_{2}\cdots\psi_{N}}+{\cal O}(\delta^{2}).
	\end{align}
	Thus, by applying this pulse sequence after the exposure to the target
    magnetic field, the state evolves to
	\begin{align}
	\ket{\Psi^{\rm CP}_{3,\delta}}=&\frac{1}{\sqrt{2}}e^{i \frac{N\Omega t}{2}}\ket{g}_{c}\otimes\ket{\psi_{1}\psi_{2}\cdots \psi_{N}}\nonumber\\
    &+\frac{1}{\sqrt{2}}e^{-i \frac{N\Omega t}{2}}\ket{e}_{c}\otimes\ket{\psi''_{1}\psi''_{2}\cdots \psi''_{N}}+{\cal O}(\delta^{2})
    \label{eq:cp_final}
	\end{align}
	We measure the controllable spin using the basis $\{\ket{\pm_{y}}=\left(\ket{e}_{c}\pm i \ket{g}_{c}\right)/\sqrt{2}\}$ and the probability of obtaining $\ket{+_{y}}$ is given by
	\begin{equation}
	P_{+y}=\frac{1}{2}+\frac{N\Omega t}{2}+{\cal O}(\delta^{2}),
    \label{eq:cp}
	\end{equation}
    Within this perturbative regime, the systematic error described in Eq.~(\ref{eq:sys_err}) is eliminated.
    An analytical evaluation for contributions of higher-order terms is provided in Appendix~\ref{app:expansion}.

    \section{Performance of the proposed pulse sequence}
    \label{sec:4}

    Here, we evaluate the performance of the proposed composite pulse sequence.
    To provide a fair comparison between the composite-pulse protocol and the conventional one reported in Ref.~\cite{PhysRevLett.115.170801} under detuning, we carefully take into account the time budgets of each protocol.

    We fix the total time $T$ allowed for the estimation of the target magnetic field.
    The number of trials $M$ is given as $M=T/\tau$, where $\tau$ denotes the duration of a single trial that contains state preparation, exposure, and readout.
    For both protocols, a larger $\tau$ ``theoretically" provides a better result, and thus $\tau=T$ is the best choice.
    However, this assumption is unrealistic because, in practice, the available time $\tau$ is always constrained by decoherence.
    Therefore, even if the probe is ideally isolated from the environment, 
    we assume that each run is limited to at most the coherence time of the probe system, 
    and we set $\tau$ to this coherence time. 
    Such an approach has also been discussed in previous research~\cite{PhysRevA.94.052320}.

    The time for one trial $\tau$ needs to be apportioned among the state preparation $t_{\rm sp}$, exposure $t_{\rm ex}$, and readout $t_{\rm ro}$.
    Note that, for the sake of clarity, we use $t_{\rm ex}$ here to represent the exposure time, replacing the notation $t$ used in previous sections.
    Let the maximum pulse strength be $\lambda_{\rm max}=1$.
    In the conventional protocol, it is desirable to allocate as long an exposure time as possible.
    Hence, pulses with the maximum strength $\lambda_{\rm max}=1$ are applied during both state preparation and readout, so that these processes are completed in the shortest possible time.
    The exposure time for the conventional protocol is then $t^{\rm conv}_{\rm ex}=\tau-4\pi$ because a $\pi$-pulse with strength $\lambda_{\rm max}=1$ requires a duration of $2\pi$ for each of the state preparation $t_{\rm sp}^{\rm conv}$ and the readout $t_{\rm ro}^{\rm conv}$.
    
    In the proposed composite-pulse protocol, the durations of state preparation and readout become comparable to the exposure duration. Hence, it takes more time to implement this pulse sequence than the conventional protocol.
    According to the pulse sequence, the duration for state preparation $t^{\rm CP}_{\rm sp}$, exposure $t^{\rm CP}_{\rm ex}$, and readout $t^{\rm CP}_{ro}$ need to satisfy the following ratio relation:
    \begin{equation}
        t^{\rm CP}_{\rm sp}:t^{\rm CP}_{\rm ex}:t^{\rm CP}_{\rm ro}=10\pi:2(8-\pi):10\pi
    \end{equation}
    where we set the maximum non-dimensional exposure time to be $2(8-\pi)$.
    To apportion $\tau$ among them keeping this ratio, we use pulses weaker than $\lambda_{\rm max}$, whose strength is
    \begin{equation}
        \lambda_{\rm CP}=\frac{20\pi+2(8-\pi)}{\tau}.
    \end{equation}
    Here, $\tau$ must be longer than $20\pi +2(8-\pi)$ to implement the composite pulse; note that the maximum pulse strength is set to $\lambda_{\rm max}=1$.
    Thus, the ratio of $t^{\rm CP}_{\rm ex}$ and $t^{\rm conv}_{\rm ex}$ is 
    \begin{equation}
        \eta:= t^{\rm CP}_{\rm ex}/t^{\rm conv}_{\rm ex}=\frac{\tau \frac{2(8-\pi)}{20\pi+2(8-\pi)}}{\tau-4\pi}.
        \label{eq:comp}
    \end{equation}
    We compare the performance of both protocols under a fixed duration $\tau$. 
    In the following analysis, we assume that the coupling strength $g$ is sufficiently large such that the rotating-wave approximation is valid.
    
    \subsection{Analytical Trade-off Analysis}
    Before presenting the numerical results, we analytically estimate the condition under which the proposed composite-pulse protocol outperforms the conventional one.
    The performance is quantified by the MSE of the estimator $\tilde{\Omega}$, which consists of the statistical variance and the squared bias:
    \begin{equation}
    \mathrm{MSE}(\tilde{\Omega}) = \mathrm{Var}(\tilde{\Omega}) + [b(\Omega)]^2.
    \end{equation}
    As discussed in Eq.~\eqref{eq:stddev_iid}, for GHZ sensing with $N$ spins and exposure time $t_{\text{ex}}$, the statistical variance scales as
    \begin{equation}
    \mathrm{Var}(\tilde{\Omega}) \approx \frac{1}{M (N t_{\text{ex}})^2},
    \end{equation}
    where we approximated $P(1-P) \approx 1/4$ for the maximum variance case.
    
    In the conventional protocol (represented by 'conv'), the exposure time $t^{\rm conv}_{\rm ex}$ is maximized within the coherence time $\tau$.
    However, it suffers from a systematic bias linear in detuning $\delta$.
    According to Eq.~\eqref{eq:mse_error}, the bias in the conventional protocol is approximately $\delta$ itself when $t^{\rm conv}_{\rm ex}$ is sufficiently large.
    Thus, the MSE is
    \begin{equation}
    \mathrm{MSE}_{\text{conv}} \approx \frac{1}{M (N t^{\rm conv}_{\rm ex})^2} + \delta^2.
    \end{equation}
    In contrast, the composite-pulse protocol (denoted by 'CP') suppresses the bias to $\mathcal{O}(\delta^2)$, which we assume to be negligible for this comparison.
    However, the composite-pulse sequence consumes a significant portion of the coherence time, reducing the available exposure time to $t^{\rm CP}_{\rm ex} < t^{\rm conv}_{\rm ex}$ (Eq.~\eqref{eq:comp}). The MSE is given by
    \begin{equation}
    \mathrm{MSE}_{\text{CP}} \approx \frac{1}{M (N t^{\rm CP}_{\rm ex})^2}.
    \end{equation}

    The composite-pulse protocol provides a better estimate when $\mathrm{MSE}_{\text{CP}} < \mathrm{MSE}_{\text{conv}}$, which leads to the following condition:
    \begin{equation}
    \frac{1}{M (N t^{\rm CP}_{\rm ex})^2} < \frac{1}{M (N t^{\rm conv}_{\rm ex})^2} + \delta^2.
    \end{equation}
    Solving for $N$, we obtain the critical number of spins beyond which the proposed method becomes advantageous:
    \begin{equation}
    N > \frac{1}{\delta t^{\rm conv}_{\rm ex} \sqrt{M}} \sqrt{\frac{1}{\eta^2} - 1},
    \label{eq:crossover}
    \end{equation}
    which reveals the trade-off between the compensation for bias and increase in the statistical error represented by $\eta$.
    In this regime, the systematic bias dominates the error budget of the conventional protocol, making the bias-cancellation capability of the composite pulse essential for achieving higher sensitivity.

    \subsection{Numerical Evaluation}    

    \begin{figure}[h]
        \includegraphics[width=80mm,page=1]{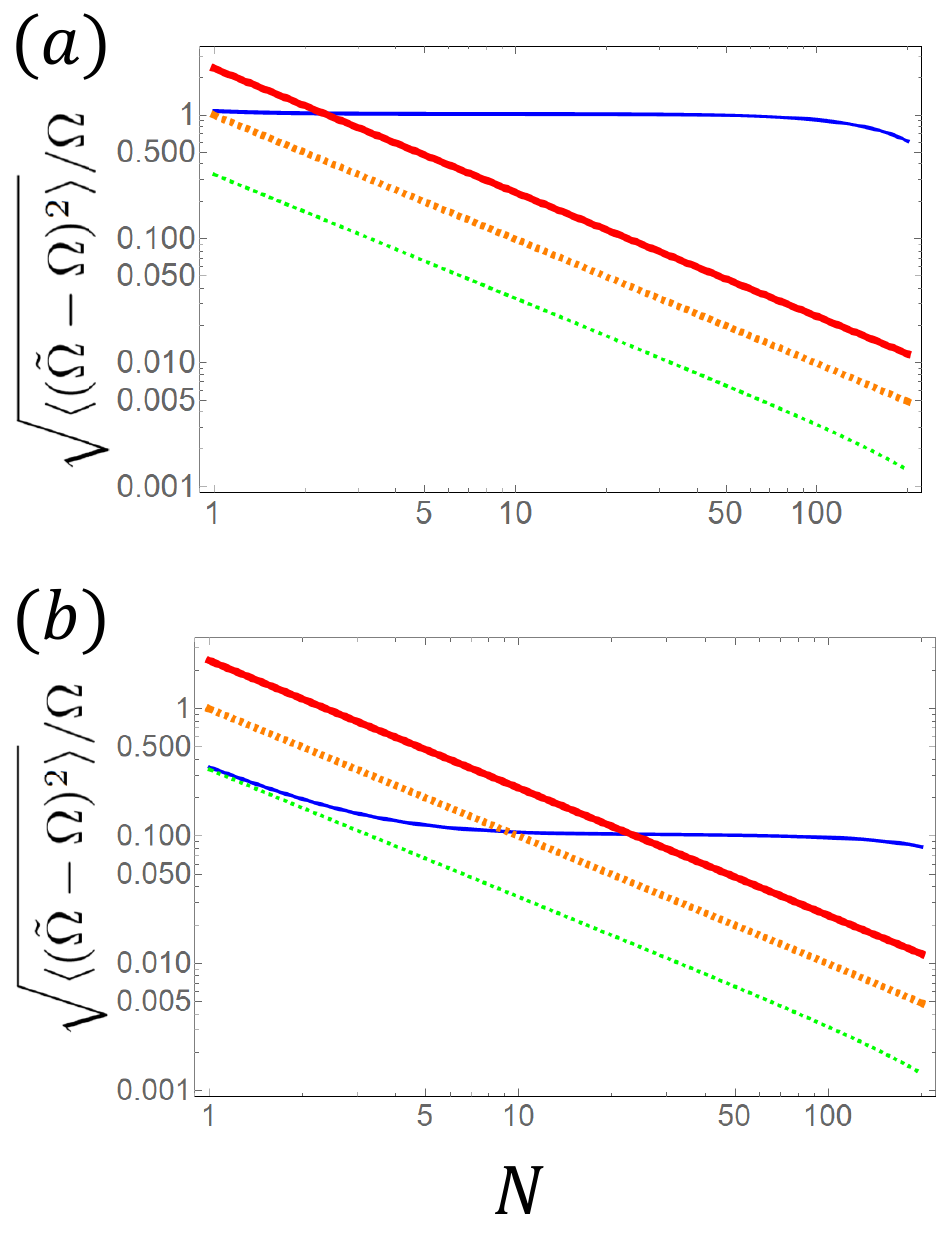}
        \caption{The relative uncertainty $\epsilon$ with respect to the number of spins $N$.
        In both panels, the blue thin line represents the relative uncertainty of the normal protocol under detuning while the red thick line represents the relative uncertainty for the case where we employ the composite pulse sequence to prepare the GHZ state.
        The orange dashed thick line indicates the relative uncertainty for the case where we use a composite pulse with changeable pulse strength, which is explained in Appendix~\ref{sec:app}.
        The green dashed line shows the Heisenberg scaling as a reference.
        We set the parameters as $M=T/\tau=10^{6}$, $\tau=100\pi$, $\Omega=0.00001$.
        The detuning is set to $\delta_{i}=\delta=~(a)~0.00001=\Omega,~(b)~0.000001=0.1\Omega$ for all $i$.
        All parameters are normalized by $\lambda_{\rm max}$.}
        \label{fig:homogeneous}
    \end{figure}       
    First, we consider the simplest case, in which detuning is homogeneous among the spins.
    we evaluate the performance of the protocol using the relative uncertainty $\epsilon$, defined as the root-mean-square error normalized by the true value:
    \begin{equation}
        \epsilon = \sqrt{\langle(\tilde{\Omega}-\Omega)^{2}\rangle}/\Omega.
    \end{equation}
    Figure~\ref{fig:homogeneous} shows the relative uncertainty of the GHZ sensing schemes with and without the composite pulse sequence.
    As shown in Fig.~\ref{fig:homogeneous}, the relative uncertainty of the conventional protocol under detuning is even far from the SQL for both cases of $(a)~\delta=\Omega$ and $(b)~\delta=0.1\Omega$.
    The performance of the conventional protocol initially follows the Heisenberg scaling, but eventually deviates from it.
    Meanwhile, the composite-pulse protocol achieves a performance close to the Heisenberg scaling ${\cal O}(N^{-1})$ over a broader region than the conventional one, albeit with an offset caused by the longer durations of state preparation and readout. It should be noted, however, that even when we employ the composite pulse, the scaling ultimately deviates from the Heisenberg limit, owing to higher-order detuning effects.
    In Fig.~\ref{fig:homogeneous}(a), we observe a significant decrease in the uncertainty of the conventional protocol around $N=200$. 
    This behavior arises from higher-order contributions of detuning.
    This leads to an oscillatory behavior of the relative uncertainty in the large-$N$ regime. 
    As a result, periodic recurrences to the Heisenberg scaling appear. 
    However, we do not exploit these points as their locations depend on the value of detuning, which is unknown.
    Up to this point, we have assumed that the pulse strength remains fixed throughout the protocol.
    However, if a change in the pulse strength during the protocol is allowed, the relative uncertainty can be improved by employing a simple composite pulse 
    (see Appendix~\ref{sec:app} for details).

\begin{figure}[t]
        \includegraphics[width=80mm,page=2]{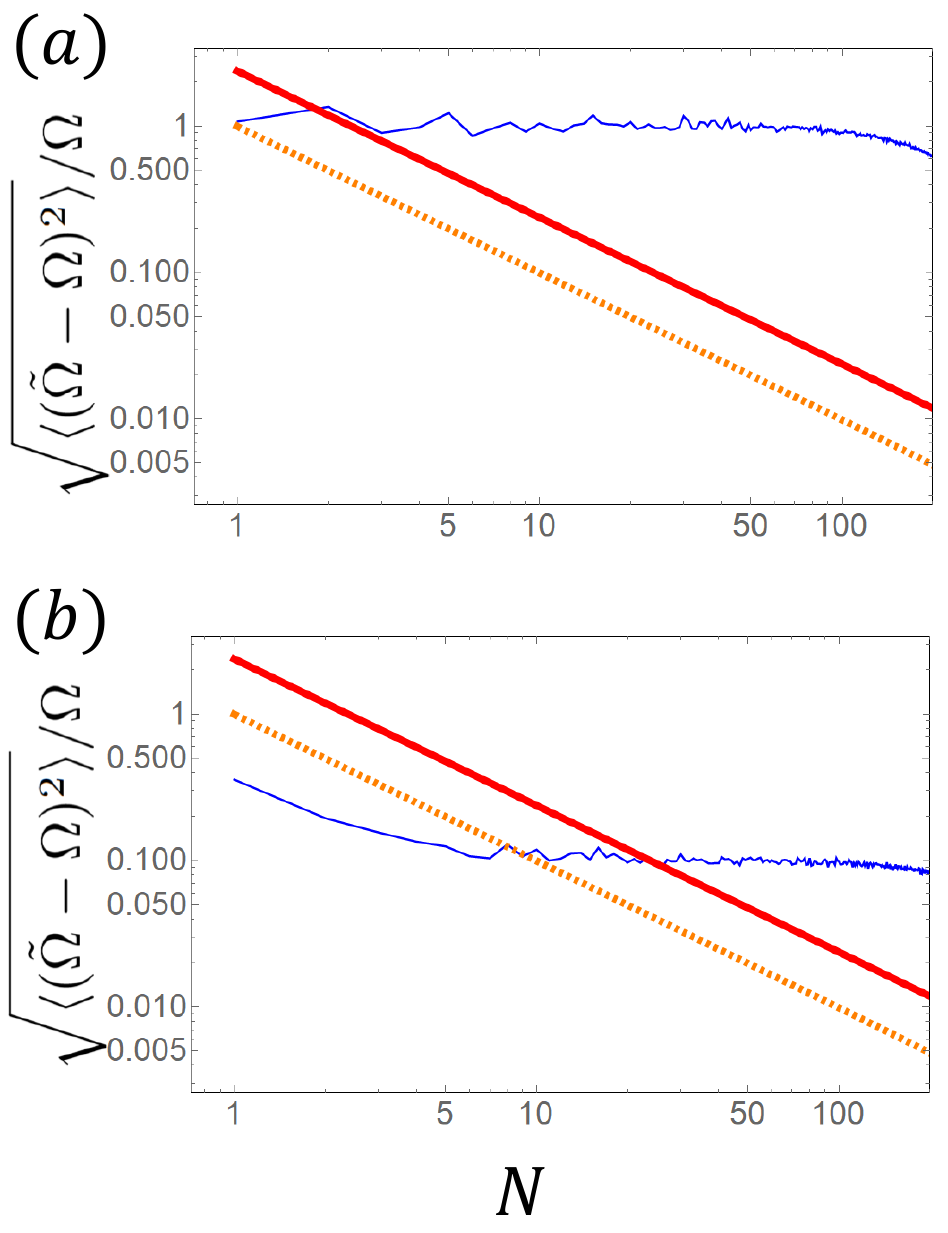}
        \caption{The relative uncertainty $\epsilon$ with respect to the number of spins $N$.
        In both panels, the blue thin line represents the relative uncertainty of the normal protocol under detuning while the red thick line represents the relative uncertainty for the case where we employ the composite pulse sequence to prepare the GHZ state.
        The orange dashed thick line is the relative uncertainty for the case where we use a composite pulse with changeable pulse strength, which is explained in Appendix~\ref{sec:app}.      
        The parameters are set to $M=T/\tau=10^{6}$, $\tau=100\pi$, $\Omega=0.00001$.
        The detuning $\delta_{i}$ is randomly sampled from the intervals: $(a)~[0,\Omega]=[0,0.00001],~(b)~[0,0.1\Omega]=[0,0.000001]$ for all $i$.
        All parameters are normalized by $\lambda_{\rm max}$.        
	}
        \label{fig:sp_inhomogeneous}
    \end{figure}

    Let us now consider a more realistic case in which detuning varies among the spins. 
    Figure~\ref{fig:sp_inhomogeneous} presents the results obtained with the same parameters as in Fig.~\ref{fig:homogeneous}, except for the detuning $\delta_{i}$.
    The detuning $\delta_{i}$ is randomly sampled from the following intervals: (a)~$[0,\Omega]$, and (b)~$[0,0.1\Omega]$.
    Even in these cases, the composite-pulse protocol yields uncertainty close to the Heisenberg scaling over a wider region than the conventional protocol. 
    We also observe that the composite pulses suppress fluctuations caused by detuning, resulting in more stable sensing performance.

    \section{Discussion and Summary}
    \label{sec:5}
    
    In this paper, we evaluated the effect of unknown detuning error on the sensing using Greenberger–Horne–Zeilinger(GHZ) states.
    While ideal GHZ sensing achieves the Heisenberg limit beyond the standard quantum limit, the resonance frequency of each spin may slightly deviate from its ideal value in practice, leading to detuning errors. We have shown that detuning during state control prevents GHZ sensing from reaching the Heisenberg limit. The resulting precision scaling could fall significantly below even the standard quantum limit, posing a serious obstacle to accurate sensing. To address this issue, we proposed a composite pulse sequence for controlling the GHZ state, which compensates for the first-order effect of detuning. Our approach improves the sensitivity under the effect of detuning, and remains effective even when detuning varies across spins.

    We employed a frequency-selective pulse to generate the GHZ state.
    Alternatively, one may generate the GHZ state directly by the coupling between the controllable spin and the memory spins, without a frequency-selective pulse.
    This can be faster than applying the pulse because we assumed in this paper that the pulse strength is sufficiently weaker than the coupling strength.
    In this case, deviation in the coupling strength can cause systematic error.
    It will be worthwhile to analyze its impact and whether we can utilize composite pulses to mitigate this deviation, such as the pulse sequence proposed in Ref.~\cite{PhysRevA.87.022323}.

    We remark on the optimality of GHZ states for sensing.
    In this work, we focused on single-parameter estimation using GHZ states, assuming the direction of the magnetic field is known a priori.
    However, in a more general setting where both the magnitude and direction of the field are unknown, the problem becomes multi-parameter estimation, for which GHZ states are not necessarily optimal~\cite{bouchard2017quantum, PhysRevA.95.052125, martin2020optimal, goldberg2021rotation}. Furthermore, while GHZ states achieve the Heisenberg limit in the asymptotic limit of many repetitions, optimal states for estimation with finite repetitions can differ from it as discussed in Ref.~\cite{PhysRevLett.85.5098}. 

    Although our current analysis is centered on the standard GHZ protocol, the issue of systematic errors arising from control imperfections, such as detuning, is ubiquitous.
    We expect that the principle of our mitigation strategy---using composite pulses to cancel coherent errors---can be extended to the preparation and readout of these more complex optimal states.
    Investigating the robustness of multi-parameter and finite-repetition sensing protocols against systematic control errors, and applying composite pulses to mitigate them, represents an intriguing direction for future research.

    Another crucial aspect for experimental implementation is the certification of the generated quantum state.
    Even with robust control sequences, verifying that the prepared state is indeed close to the target GHZ state is essential for guaranteeing sensing precision.
    Recent studies have proposed feasible schemes for state verification, such as the efficient protocols developed for multipartite entanglement~\cite{martins2024experimental, Schmid:24}.
    Integrating these certification methods with our composite-pulse protocol would allow for a complete validation of the sensing process—confirming that the systematic errors are suppressed and high-fidelity GHZ states are successfully produced even under detuning.

    Finally, we discuss the relationship between our method and recently developed quantum error mitigation (QEM) techniques for metrology~\cite{PhysRevLett.129.250503, PhysRevA.109.022410, kwon2025virtual,lin2025error}.
    QEM approaches aim to reduce systematic bias primarily through post-processing of measurement data.
    However, most QEM approaches in metrology focus on mitigating bias caused by the parameter uncertainty of stochastic noise channels.
    To apply standard QEM protocols effectively to coherent errors like detuning, one typically employs techniques such as Pauli twirling (randomized compiling)~\cite{PhysRevA.94.052325} to convert coherent errors into stochastic Pauli noise.
    However, this strategy introduces two types of overheads: the additional gate operations required for twirling, which may introduce further errors, and the sampling overhead inherent to QEM, where the statistical variance increases to compensate for the bias.
    In contrast, our composite-pulse approach suppresses systematic bias at the physical Hamiltonian level, prior to the measurement process.
    This "active" cancellation ensures that the raw data itself is high-quality, avoiding both the hardware complexity of twirling and the sampling overhead of probabilistic error cancellation.
    Crucially, these paradigms can be complementary.
    Applying our composite-pulse protocol to suppress the leading-order coherent errors significantly reduces the error rate, thereby 
    lowering
    the cost of any subsequent QEM protocols required to eliminate residual higher-order errors. 
    
    \section*{Acknowledgment}
    This project is supported by
    JST Moonshot R\&D Grant
    Number JPMJMS226C, 
    JST CREST Grant Number JPMJCR23I5, and Presto
    JST Grant Number JPMJPR245B.

    \appendix
    
    \section{Validity of Approximations}
    \label{app:approximations}
    In this appendix, we quantitatively justify the two major approximations employed in our analysis: the Rotating Wave Approximation (RWA) and the perturbative treatment of detuning.

    \subsection{Rotating Wave Approximation}
    \label{app:RWA}
    In the main text, the time evolution of the system during pulses is described by the Hamiltonian in Eqs.~\eqref{eq:decomposed_hamiltonian} and \eqref{eq:erroneous2}.
    When applying a pulse to the state $\ket{\Psi_0}$ in Eq.~\eqref{eq:initial} for example,
    we employ the effective Hamiltonians in Eq.~\eqref{eq:quater}.
    These Hamiltonians are derived under the RWA.
    To justify this, we trace the derivation of these Hamiltonians from the full Hamiltonian and evaluate the impact of the terms neglected in the RWA.
    
    For simplicity, we treat the approximation of Eq.~\eqref{eq:decomposed_hamiltonian} by Eq.~\eqref{eq:quater} when the pulse frequency $\omega$ is set to $\omega_{+}$, but the same discussion is valid for the other cases.
    The full Hamiltonian~\eqref{eq:decomposed_hamiltonian} is rewritten as
    \begin{equation}
    \tilde{H}(t) = \ket{g}_{c}\bra{g} \otimes V^{\text{CRT}}_{-} + \ket{e}_{c}\bra{e} \otimes (\tilde{H}_{\rm res} + V^{\text{CRT}}_{+}),
    \end{equation}
    where $\tilde{H}_{\rm res}$ is given in Eq.~\eqref{eq:quater}, and $V^{\text{CRT}}_{\pm}$ represents the counter-rotating terms:
    \begin{align}
    V^{\text{CRT}}_{+} =& \frac{1}{4} \left( J^{(m)}_{x}\cos(2\omega_{+} t - \phi) + J^{(m)}_{y}\sin(2\omega_{+} t - \phi)\right),\nonumber\\
    V^{\text{CRT}}_{-} =& \frac{1}{4} \left( J^{(m)}_{x}\cos(2 \omega t - \phi) + J^{(m)}_{y}\sin(g t - \phi) \right)
    \end{align}
    The validity of replacing $\tilde{H}(t)$ with Eq.~\eqref{eq:quater} is determined by the ratio of the pulse strength to the driving frequency.
    According to time-dependent perturbation theory, the correction from $V^{\text{CRT}}_{\pm}$ (such as the Bloch-Siegert shift) scales as $\mathcal{O}(1/\omega_{+})$, $\mathcal{O}(1/\omega)$ and $\mathcal{O}(1/g)$, where all parameters are normalized by $\lambda$.
    Consequently, the error introduced by the RWA is strictly bounded by these orders.
    When $\omega$ and $g$ are sufficiently large ($\gg 1$), these RWA errors are significantly smaller than the errors caused by detuning $\delta$, which we aim to mitigate.

    \subsection{Validity of Perturbative Expansion}
    \label{app:expansion}

    We justify the neglect of higher-order terms of the detuning $\delta$ in Eq.~\eqref{eq:cp}.
    First, we rewrite the final state $\ket{\Psi^{\rm CP}_{3,\delta}}$ generated by the protocol as
    \begin{align}
    \ket{\Psi^{\rm CP}_{3,\delta}} = \frac{1}{\sqrt{2}}\ket{g}_{c}\otimes\ket{\chi_{1}\cdots \chi_{N}}
    +\frac{1}{\sqrt{2}}\ket{e}_{c}\otimes\ket{\chi''_{1}\cdots \chi''_{N}},
    \end{align}
    where $\ket{\chi_{i}}$ and $\ket{\chi''_i}$ are the states of the $i$-th memory spin.
    Contributions of $\Omega$ and higher-order terms of $\delta$ are incorporated into these states, unlike the case of $\ket{\psi_{i}}$ and $\ket{\psi''_{i}}$ in the main text.
    We rewrite $P_{+y}$ in the main text as $P(\Omega,\vec{\delta})$, and it is given by
    \begin{equation}
    P(\Omega,\delta_{1},\cdots,\delta_{N}) = \frac{1}{2} + \frac{1}{2}\text{Im}\left(\prod^{N}_{i=1}f_i(\Omega,\delta_{i})\right),
    \end{equation}
    where $f_i(\Omega,\delta_{i}) = \braket{\chi_{i}}{\chi''_{i}}$.
    Assuming uniform detuning $\delta_{i}=\delta$, this reduces to $P(\Omega,\delta) = \frac{1}{2} + \frac{1}{2}\text{Im}(f(\Omega,\delta)^N)$.
    Although this is the simplest case, the following discussion is straightforwardly extended to the case of different $\delta$.

    To analyze the contribution of higher-order terms, we parameterize the dependence on $\Omega$ and $\delta$ using a scalar variable $s \in [0, 1]$:
    \begin{equation}
    g(s) := P(s\Omega, s\delta).
    \end{equation}
    According to Taylor's theorem, there exists $0 \le \zeta \le 1$ that satisfies the following equality:
    \begin{equation}
    P(\Omega, \delta) = g(1) = g(0) + g'(0) + \frac{1}{2} g''(\zeta),
    \end{equation}
    where the prime symbol denotes differentiation with respect to $s$.
    As our protocol cancels the first-order error with respect to $\delta$,
    and the first-order dependence on $\Omega$ yields the signal, we have $g'(0) = N\Omega t/2$.
    The limitation of our approximation corresponds to the Lagrange remainder term $g''(\zeta)$.

    We bound $g''(\zeta)$ by evaluating the derivatives of the unitary operators.
    Using the product rule and the triangle inequality, the remainder term is bounded for any $s$ by
    \begin{equation}
        |g''(s)|\leq N |f''(s)| |f(s)|^{N-1}+N(N-1)|f'(s)|^{2}|f(s)|^{N-2},
        \label{eq:chain}
    \end{equation}
    where $f(s):=f(s\Omega,s\delta)$.
    Note that the function $f(1)$ is the inner product $\braket{\chi_{i}}{\chi''_{i}}$, and therefore $f(s)$ is rewritten as
    \begin{equation}
        f(s)=\bra{g} U^{\dagger}_{g}(s) U_{e}(s)\ket{g}
    \end{equation}
    where $U_{g,e}$ is the whole dynamics with the parameters $(s\Omega,s\delta)$ of each memory qubit when it couples with the ground and excited states of the control qubit, respectively.
    The right-hand side is bounded from above by the operator norm (denoted by $||\bullet||_{\rm op}$) as
    \begin{equation}
    |\bra{g} U^{\dagger}_{g}(s) U_{e}(s)\ket{g}| \leq ||U^{\dagger}_{g}(s) U_{e}(s)||_{\rm op}=1,
    \end{equation}
    Note that the operator norm of unitary operators is one.

    The derivatives of $f(s)$ satisfy
    \begin{align}
    |f'(s)| \leq& ||\left(U^{\dagger}_{g}(s)\right)'||_{\rm op}+||\left(U_{e}(s)\right)'||_{\rm op}\nonumber\\
    |f''(s)| \leq& ||\left(U^{\dagger}_{g}(s)\right)''||_{\rm op}+||\left(U_{e}(s)\right)''||_{\rm op}\nonumber\\
    &+2||\left(U^{\dagger}_{g}(s)\right)'||_{\rm op}||\left(U_{e}(s)\right)'||_{\rm op}
    \label{eq:op_norm}
    \end{align}
    where we use $||AB||_{\rm op}\leq ||A||_{\rm op}||B||_{\rm op}$ for any matrices $A$ and $B$.
    We explicitly derive the bounds for the derivatives of the unitary operators $U_{\alpha}(s)$ ($\alpha\in \{g,e\}$).
    The unitary operators $U_{\alpha}$ are schematically given by 
    \begin{equation}
        U_{\alpha}(s)={\cal T}\exp\left(-i \int^{\tau}_{0}\left(H^{\rm CP}_{\alpha}(t)+s H^{\Omega,\delta}(t)\right)d t\right),
    \end{equation}
    where ${\cal T}$ represents the time-ordered product and $\tau$ the duration of a single trial.
    The first term $H^{\rm CP}_{\alpha}(t)$ represents the composite-pulse Hamiltonian during the state preparation and readout procedures, and $H^{\rm CP}_{\alpha}(t)=0$ during exposure.
    The second term $H^{\Omega,\delta}(t)$ represents the effects of $\Omega$ and $\delta$:
    \begin{equation}
        H^{\Omega,\delta}(t)=
        \begin{cases}
            \delta\frac{\sigma_{z}}{2},~~\text{(during state preparation and readout)}\\
            (\Omega + \delta)\frac{\sigma_{z}}{2}.~~\text{(during exposure)}
        \end{cases}
    \end{equation}
    The derivatives of the unitary operator $U_\alpha(s)$ satisfy
    \begin{equation}
    \left\| \frac{d^k U_\alpha}{ds^k} \right\|_{\rm op} \le \left( \int_0^\tau \left\| H^{\Omega,\delta}(t) \right\|_{\rm op} dt \right)^k \le \left[ \frac{1}{2}(|\Omega|+|\delta|)\tau \right]^k.
    \label{eq:duha}
    \end{equation}
    Combining Eqs.~\eqref{eq:chain}, \eqref{eq:op_norm} and \eqref{eq:duha}, we obtain the upper bound of the remainder term $g''(\zeta)$ as
    \begin{equation}
        |g''(\zeta)|\leq  \left[N\tau \left(|\Omega|+|\delta|\right)\right]^{2}.
    \end{equation}
    for any $\zeta$.
    Hence, our approximation of neglecting higher-order terms of $\Omega$ and $\delta$ is valid when $|\Omega|+|\delta|\ll 1/N\tau$.
    
    Although the above bound is exact, the actual contribution of higher-order terms is likely smaller.
    Note that in $f(\Omega, \delta)$, the variables always appear in the form of $i\Omega$ and $i\delta$, which is a basic property of quantum dynamics.
    This implies that the function $f(s\Omega, s\delta)$ contains imaginary coefficients for odd powers of $s$ and real coefficients for even powers of $s$.
    As $P$ depends on $\text{Im}(f^N)$, purely real second-order terms (proportional to $s^2$, such as $\delta^2$ and $\Omega\delta$) do not contribute to $P$.
    Thus, the dominant systematic error is expected to arise from third-order terms in $\Omega$ and $\delta$, indicating a wider validity range.
    
    \color{black}

    \section{protocols when more meticulous pulse control is allowed}
    \label{sec:app}

    We analyze the scenario in which the pulse strength is tunable.
    The maximum pulse strength is normalized to $\lambda_{\rm max}=1$. 
    As we noamrlize all parameters by the maximum pulse strength, applying a pulse of strength $\lambda$ effectively rescales thedetuning as $\delta/\lambda$.
    Using this observation, we can tune the pulse strength to compensate for unknown phase shifts throughout the process.
    During the state preparation, we apply the following composite pulse with variable pulse strength:
	\begin{enumerate}[I.]
	\item $\omega=\omega_{-}$, $\phi=-\pi/2$, $\lambda=4\pi/(t+2\pi)$, $s/2=\pi/\lambda$, 
	\item $\omega=\omega_{+}$, $\phi=-\pi/2$,  $\lambda=4\pi/(t+2\pi)$, $s/2=\pi/\lambda$, 
	\item $\omega=\omega_{-}$, $\phi=-\pi/2$, $\lambda=1(=\lambda_{\rm max})$, $s/2=\pi/\lambda$, 
	\end{enumerate} 
    where $t$ is the exposure time.
    After the interaction with the magnetic fields, the readout process is performed in a manner similar to the state preparation by applying the following pulses:
	\begin{enumerate}[I.]
	\item $\omega=\omega_{-}$, $\phi=-\pi/2$, $\lambda=1$, $s/2=\pi/\lambda$, 
	\item $\omega=\omega_{+}$, $\phi=-\pi/2$, $\lambda=4\pi/(t+2\pi)$, $s/2=\pi/\lambda$, 
	\item $\omega=\omega_{-}$, $\phi=-\pi/2$, $\lambda=4\pi/(t+2\pi)$, $s/2=\pi/\lambda$.
	\end{enumerate}     
    Note that each pulse in the state preparation and readout processes is simply a $\pi$-pulse, as $s\lambda/2=\pi$ for each case.
    A straightforward calculation shows that this pulse sequence cancels the effect of detuning on the probability $P_{y+}$ up to first order.
    Consequently, we obtain the same result as in Eq.~\eqref{eq:cp}.

    \bibliography{sensing}

\end{document}